\documentclass[twocolumn,aps,pre,showpacs,amsmath,amsfonts,amssymb,floatfix]{revtex4}
\usepackage{morefloats}
\usepackage{amsmath}
\usepackage{graphicx}
\usepackage{dcolumn}
\usepackage{bm}

\begin{document}

\title{Optimization of synchronizability in multiplex 
networks by rewiring one layer}
\author{Sanjiv K. Dwivedi$^{1}$, Murilo S. Baptista$^{2}$ and Sarika Jalan$^{1,3}$ }
\affiliation{1. Complex Systems Lab, Physics Discipline, Indian Institute of Technology Indore, Khandwa Road, Indore 452017, India}
\affiliation{2. Institute of Complex Systems and Mathematical Biology, SUPA, University of Aberdeen, UK}
\affiliation{3. Center for Biosciences and Biomedical Engineering, Indian Institute of 
Technology Indore, Khandwa Road, Indore 452017, India}

\begin{abstract}
The mathematical framework of multiplex networks has been
increasingly realized as a 
more suitable framework
for modelling real-world complex systems.
In this work, we investigate the optimization of synchronizability in multiplex networks
by evolving only one layer while keeping other layers fixed.
Our main finding is to show the conditions under which the efficiency of convergence to the most optimal 
structure is almost as good as the case where both layers are rewired during an 
optimization process. In particular, inter-layer coupling strength responsible for the integration 
between the layers turns out to be crucial factor governing the efficiency of optimization even for the cases when the layer
going through the evolution has nodes interacting much weakly than those
in the fixed layer.
Additionally, we investigate the dependency of synchronizability on the 
rewiring probability which governs the network structure from a regular lattice to the 
random networks.
The efficiency of the optimization process preceding evolution driven by the optimization process
is maximum when the fixed layer has regular architecture,
whereas
the optimized network is more synchronizable for the fixed layer
having the rewiring probability lying
between the small-world transition and the random structure. 
\end{abstract}

\maketitle 

{\it Introduction:}
The framework for a single network has been extremely successful for predicting and understanding
behaviour of complex systems
\cite{net_rev}.
However, recent studies of multiplex 
networks are providing  new insights 
to the research in real-world complex 
systems by incorporating in the analysis
the fact that they are composed by 
several types of networks (layers) and 
more than
one type of interactions exist among 
the layers.
Thus, multiplex networks are expected to provide better understanding about the 
underlying structural and dynamical properties of real-world 
systems as compared to the traditional 
isolated networks approach \cite{Boccaletti2}.
For instance, diffusion
processes taking place on multiplex networks has been shown to exhibit abrupt transitional behaviour 
guided by inter-layer coupling strength \cite{Radicchi}.
Entropy rates and information transmission was shown to be strongly
regulated by the ratio between inter-connectivity and the size
of the single layer \cite{Murilo2016}. Similarly, cluster synchronization
of a layer in multiplex networks has been demonstrated to be strongly
affected by the network parameters of other layer \cite{SJ_clus_multi2016}. 
Furthermore, endemic states in multiplex networks has been shown to crucially
depend on the interconnectivity of the layers,
not emerging in individual layers when considered in isolation \cite{Saumell}.
The multiplex network framework has allowed to incorporate
various new interconnected processes into the modelling of complex systems,
such as the work in Ref.~\cite{Granell} that studies 
the spreading of an epidemic 
in
individuals 
contributing to the understanding of how
disease spreading can be controlled.

Further, synchronization phenomena or collective behaviour of coupled dynamical units
has been a topic of intensive research \cite{book_kurths}.
Dynamical behaviour of interacting units depends on the structural properties of interactions. One such relation between the structural property of a network
and the synchronous dynamical behaviour of units interacting via 
diffusive coupling is measured by the synchronizability of the 
network, defined by the ratio between the
first nonzero and the largest eigenvalues of the corresponding Laplacian 
matrix  \cite{Pecora,Barahona}.
Larger (smaller) the $R$ values, the smaller (the larger) 
coupling strength interval for which synchronization is observed.

The most optimized network in terms of synchronizability 
has been shown to exhibit homogeneity in its  degree distribution and in the
betweenness centrality of the nodes \cite{Luca}.
Optimization of synchronizability in networks with nodes
connected by weighted strengths is a problem with an extra dimension of complexity.
However, it has been shown that such networks can be successfully evolved to 
become optimally synchronizable \cite{Zhou,Chavez}.
Even
more challenging is the optimization of multiplex networks, which would require optimization strategies involving several network parameters and larger dimensional systems. Take the brain as an example, it learns by rewiring its synaptic connections. If the brain were to adapt (optimize behavior) based on all its possible scenarios, that would be a fantastic complex optimization process. Rather, it is plausible to think that optimization in the brain (such as those driven by Hebbian learning rules) is driven by evolution rules applied locally.  This paper shows that indeed  synchronizability of a whole multiplex network can be achieved by rewiring only one layer, thus showing that the computational complexity of optimization in multiplex networks can be drastically reduced.  

More specifically, we study optimization of a layer in multiplex network
such that the entire network becomes more synchronizable. During the evolution, only one layer
is rewired while keeping the other layer(s)'s topology fixed. 
Changing the network architecture of one layer affects the 
dynamical evolution of the other layers because of the interactions
mediated by the inter-layer couplings. We therefore investigate
the efficiency of the optimization in terms of the interplay 
between the intra-layer coupling strengths of the layer going 
through the evolution process and inter-layer couplings. 
Furthermore, we investigate the impact of the network architecture of the fixed layer on the optimization efficiency. 
Our investigation reveals that the inter-layer coupling strength plays a crucial role
in determining the impact of the optimization process on the 
synchronization of the entire
network. Interestingly, even if the layer going through the evolution has much weaker
intra-layer coupling strength as compared to that of the fixed layer, efficiency of
optimization is high if there is a strong interaction between
the layers.
Moreover, the optimization leads to the best synchronizable multiplex network
when the network architecture of the fixed layer lies
between a complete random architecture and the one observed at the small-world transition arising
due to the combined impact of the degree homogeneity and the diameter. 

Optimization of complex networks is behind the success of technological as 
well as natural adaptive processes. The brain learns by rewiring its synaptic 
connections. Deep learning machines changes internal structures of its 
neural network to optimize its logical outputs. It is a current 
scientific challenge to understand natural optimization processes in order to 
reproduce it. The difficulty lies on the fact that optimization complexity 
increases exponentially by the size of the system. This paper shows that 
synchronizability of a whole multiplex network, the ability of the network to 
synchronize, can be optimized by only rewiring a single network layer. Thus, 
this paper opens up a new avenue of research, by showing that optimization 
complexity can be drastically optimized.   

{\it Theoretical Framework:} Let $A$ and $B$ be two adjacency matrices
with dimension $N \times N$ corresponding to network configurations 
representing the initial structure of two layers of a multiplex network. 
The elements in the adjacency matrices [$a_{ij}$ and $b_{ij}$] take 
value 1 and 0 depending upon whether there exists a connection between the 
$i$ and $j$ nodes or not.  
We perform optimization for 
individual layers with several architectures.
The weighted adjacency matrix of the multiplex networks can be written as,
\begin{equation}
 M = \left[
    \begin{array}{cc}
      A~~D_{x}I\\
      D_{x}I^T~~E_{y}B
    \end{array}
\right]
\end{equation}
where $E_y$ is the intra-layer coupling strength of the layer,  $D_x$ represents the inter-layer coupling strength, and
$I$ is the inter-layer adjacency matrix representing the connections 
from $B$ to $A$, and $I^T$ (the transpose of $I$) represents the connections from
layer $A$ to $B$.  

We optimize the eigenvalue ratio $(R) = \frac{\lambda_{max}}{\lambda_{2}}$, inverse of synchronizability, where
$\lambda_{max}$ and $\lambda_{2}$ are the largest and the first non-zero eigenvalue of the 
Laplacian matrix of the multiplex network constructed from
$\sum_{j=1}^{2N} M_{ij} \mathbb{I} - M$, where $\mathbb{I}$ represents the identity matrix.
We use the simulated annealing technique \cite{ref_SA}
to perform the optimization of $R$. Our optimization aims
at minimizing $R$, and thus, maximizing synchronizability.
This optimization 
technique has several variations depending upon the problem in hand.
For the current work, the method is explained as follows. We take an 
initial multiplex network with a given set of parameters. Next, we calculate the  eigenvalue ratio $R_1$ of 
the corresponding Laplacian matrix of the initial multiplex network. Rewiring
is performed only in one layer by keeping the second layer's architecture 
fixed throughout the evolution.
We calculate the eigenvalue ratio $R_2$ of the multiplex network after performing
a single rewiring. 
The initial multiplex network is replaced by the rewired multiplex network if
the latter is more synchronizable and $R_2$ $\le$ $R_1$ otherwise replaced with the probability $p=\exp((R_1-R_2)/T)$.
Whereas, the initial network is selected with the probability $1-p$. 
$T$ is a constant taken initially 1.000. It is updated to the end of each generation by 
0.999T.

\begin{figure}[t]
\centerline{\includegraphics[width=1.0\columnwidth]{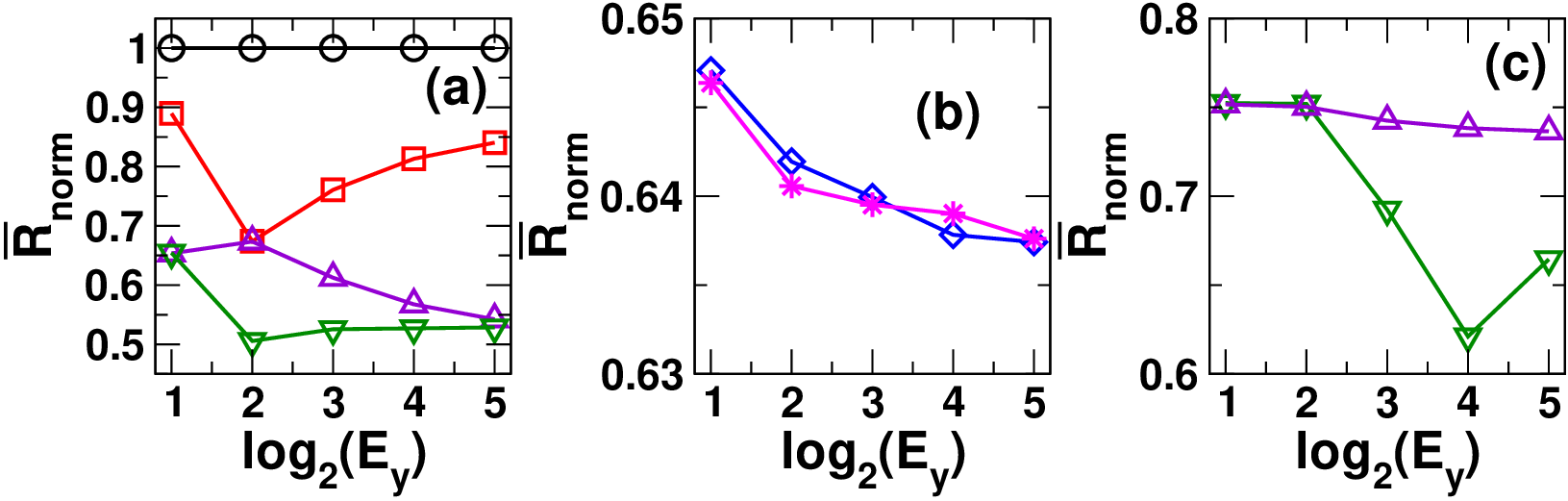}}
\caption{Picture shows $\bar{R}_{norm}$ against 
$E_y$ for several optimization  configurations. 
(a) $R_{norm}$ for case (I) when rewiring is 
performed in the layer $A$ and 
$D_x$ takes value 1 (circles), case (II) 
when rewiring is also performed in the layer A, 
but inter-layer coupling is strong, i.e. 
$D_x$ takes values from 2 to 32 (squares), 
case (III) which corresponds to the scheme 
when $D_x$ again varies from 2 to 32 
and rewiring is done in the layer having stronger
intra-layer coupling (layer $B$) 
(upper triangles). Case (IV) corresponds to the 
situation when rewiring is performed in both 
$A$ and $B$ layers with $2 \le D_x \le 32$ 
(lower triangles). (b) 
$R_{norm}$ for case (V) which is similar to the 
case (IV) except that the inter-layer coupling
is weak, i.e. $D_x$ takes value 1 
(stars) and case (VI) which corresponds to 
the rewiring performed in layer $B$ having
stronger intra-layer coupling strength 
with again $D_x$ being 1 (diamonds).
We consider $D_{x} = E_{y}$ for the cases having 
stronger inter-layer coupling strengths. 
For (a) and (b), 
$\langle k\rangle$ of each layer is 10 with $N=500$. 
(c) Behaviour of $R_{norm}$ for the case (III) 
(upper triangles) and case (IV) (lower triangles). Network parameters
are $\langle k\rangle = 20$ with size $N=500$. 
For each case, optimization minimises $R$ for 200,000 iterations.}
\label{Case}
\end{figure}
During the optimization process, the fixed layer introduces a limit to the synchronizability
of the entire multiplex network. Nevertheless, the effect of the fixed layer varies depending upon 
inter and intra-layer coupling strengths of both the layers. Naturally, if the layer going 
through the rewiring during evolution has stronger intra-layer couplings as compared to that
of the fixed layer,
the optimization should be more efficient. Interestingly, we find that the
inter-layer coupling
strength $D_x$ has more profound impact on the optimization.
To observe the impact of $E_{y}$ and $D_{x}$ on the efficiency of the 
optimization process, we systematically investigate the following cases.  In case (I), 
inter-layer coupling
strength is weak, i.e. $D_x$  takes the value  1 and the layer with weaker
intra-layer coupling strengths (i.e. layer $A$) is rewired resulting in evolution of this layer, 
whereas the architecture of the layer
with stronger intra-layer coupling strengths (layer $B$) 
is maintained throughout the 
evolution process. 
In case (II), inter-layer coupling strength is strong ($D_x$ is large), 
and other parameters are the same as for the case (I).
In case III, $D_x$ is large and the layer with smaller intra-layer coupling 
(layer $A$) is preserved during
the evolution.  The rewiring is performed only in the layer having larger 
intra-layer coupling strength (layer $B$). To compare the results about the 
impact of change 
in only one layer on the synchronizability of the entire multiplex network with
those obtained for changes in both the layers, 
we consider two more cases.  In case (IV) and (V), evolution is allowed in 
both the layers with case (IV) considering $D_{x} > 1$ and
case (V) considering $D_x=1$.  In case (VI), 
$D_x=1$ and the layer with weaker intra-layer coupling strengths (layer $A$)
is preserved, and the layer with stronger intra-layer couplings is evolved.
Further, we measure efficiency of synchronizability by 
$R_{norm} = \frac{R_{opt}}{R_{ini}}$, where $R_{opt}$ and $R_{ini}$ represent value
of $R$ for the final optimized and the initial multiplex network, respectively. 
As the eigenvalue ratio ($R$) and 
the synchronizability of a network are inversely related, the lower the $R_{norm}$ value, the
better is the efficiency of the synchronization.

{\it Results:}
\begin{figure}[t]
\centerline{\includegraphics[width=1.0\columnwidth]{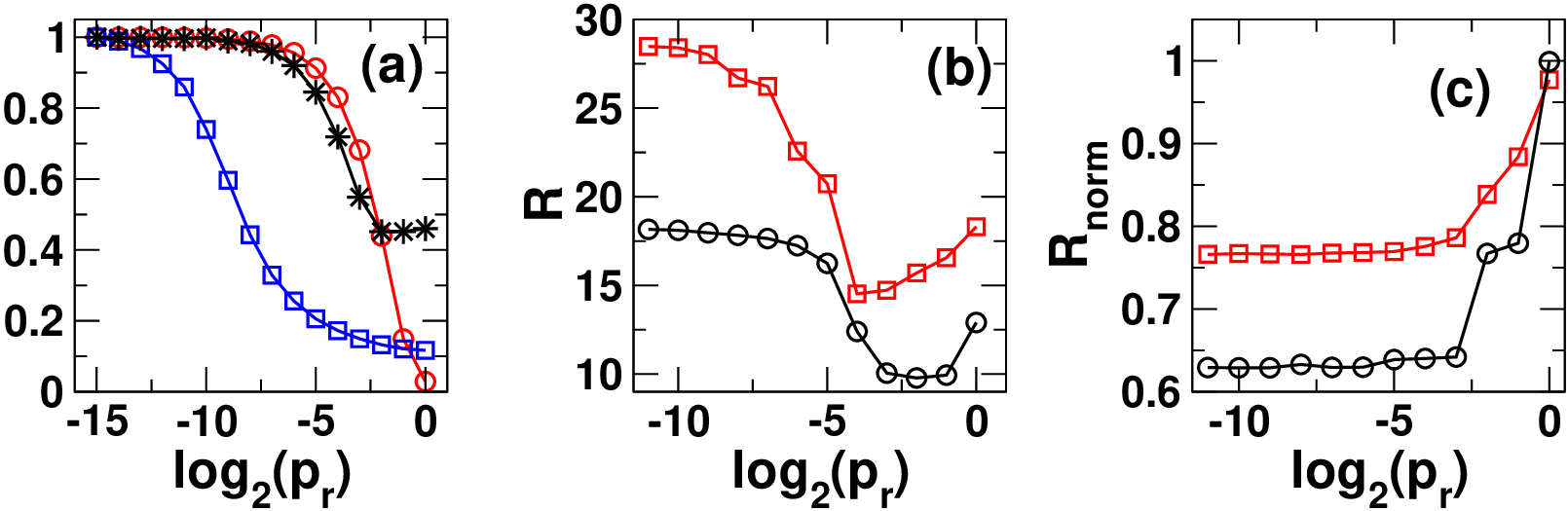}}
\caption{(a) Clustering coefficient (circles), 
characteristic path length (squares) and 
normalized eigenvalue ratio (star) as a function of 
small-world rewiring probability ($p_r$) of an initial multiplex network
having one layer represented by a small-world network with $p_r$ rewiring
probability and other layer represented by ER network.  
(b) Depicts impact of $p_r$ on 
the optimized $R$ value for the 
average degree of each layer taken as $\langle k\rangle$ = 10 (circles) 
and $\langle k \rangle$=20 (square). The fixed layer is 
represented by a
small-world network with $p_r$ probability
and the layer represented by ER network is evolved through the optimization mechanism.
(c) Shows the impact of $p_r$ on the 
optimization efficiency ($R_{\mathrm{norm}}$) for $\langle k\rangle$ = 10 (circles) and $\langle k\rangle$ = 20 (squares).
Each layer of the multiplex networks has $N_1=N_2=500$ and value
of $D_x=1$.}
\label{SWtr}
\end{figure}
As evolution progress, the optimization attempts to bring the layer going
through the rewiring to a structure which is favourable for synchronization,
whereas the fixed layer imposes a limit to the synchronizability or on
the efficiency of the synchronization. 
Fig.~(1) demonstrates that for the case (I), optimization does
not succeed in producing a synchronizable networks for any value of $E_y$ 
we have considered.
Whereas in the case (II), the optimization succeeds into finding synchronizable 
networks for all the values of $E_{y}$ considered here. Though, the maximum
efficiency corresponds to a value of $E_{y}$ for which $R_{norm}$
is minimal, the exact value
of $E_y$ for which efficiency is maximal depends on the size and average degree of the 
network. 
Further, a low value of $D_x$ typically produces a low
value of $\lambda_2$, whereas
high values of $D_x$ lead to high value of $\lambda_{max}$ \cite{Ribalta}. Both these factors contribute to an increase in the
$R$ values and for the model considered here 
$R$ can be determined as following: 
For $D_{x}$ being smaller with respect to $E_y$, referred as weaker $D_x$ case,
one can understand
the behaviour of $R$ using the following approximation:
\begin{equation}
 R \approx \frac{\max\limits_{\alpha}\big[\lambda_{max}(L^{\alpha}) ~+~D_{x}\big]}{2D_{x}}
\label{EQ2}
\end{equation}
where $L^\alpha$ is Laplacian of the $\alpha^{th}$ layer, 
$\lambda_{max} (L^\alpha)$ is maximum eigenvalue of
the Laplacian of the $\alpha^{th}$ layer. For the model considered in Eq.~(1), the
$\alpha$ index represents the matrix $A$ or matrix $E_y B$, and therefore 
$L^{\mathrm{A}} = \sum_j A_{ij} \mathbf{I} - A$, and 
$L^{B} = \sum_j E_y B_{ij} \mathbf{I} - E_y B$. 

When $D_{x} > 1$, i.e., inter-layer being stronger than the intra-layer;
\begin{equation}
 R \approx \frac{2D_x + \sqrt{2}\lambda_{max}(L^{\mathrm{AV}})}{\lambda_{2}(L^{\mathrm{AV}})}
\label{EQ3}
\end{equation}
where $L^{AV}$ is the average Laplacian of two layers.

For small $D_{x}$ values, $R$ is governed by 
Eq.~(\ref{EQ2}). Since $\lambda_{max}$ of the fixed layer having  
stronger intra-layer
coupling strength governs the numerator of Eq.~(2) which leads to
the same value of $R$ throughout the optimization 
resulting in $R_{norm} \cong 1$.
For larger $D_{x}$ values, Eq.~(\ref{EQ3}) starts to dominate over Eq.~(\ref{EQ2}).
The layer going through the evolution, even though having
smaller intra-layer couplings as compared to those of the fixed layer, contributes to $R$ as  
because of the average value of the Laplacians of both the layers 
appearing in the denominator of Eq.~(3).
Further, structural changes caused by the evolution process are capable of steering 
$\lambda_2$ of the evolved layer towards larger values, resulting in the
smaller $R$ values (Eq.~(3)) and therefore, optimization is successful.
For a further increase in $D_x$, Eq.~(\ref{EQ3}) holds even better for the 
$R$ values, and suddenly there is an increase in the efficiency of the 
optimization. However, the larger the values of $D_{x}$ and $E_{y}$ are, the
stronger the contribution of the fixed layer coupling strength in $L^{AV}$ of
Eq.(\ref{EQ3}) is. As a result, the efficiency again decreases for the 
case (II).  Efficiency for the cases (V) and (VI), i.e. for smaller 
values of  
$D_{x}$, can be explained by Eq.~(\ref{EQ2}) where $\lambda_{max}$ comes from the 
rewired layer, which has stronger intra-layer couplings and hence always dominates
the numerator of Eq.~(3).
Interestingly, for smaller $D_x$ values, rewiring in both the layers (case (V)) does not lead to an increase in the efficiency 
as compared to the rewiring in a single layer having stronger coupling strength (case (VI))
as illustrated in (Fig.~1(b)). For 
larger values of $D_{x}$ and $E_{y}$, Eq.~(\ref{EQ3}) 
controls the values of $R$ where structural properties of both the layers are crucial to 
determine the spectral properties of the $L^{AV}$ matrices.
As a result, the efficiency is higher for the case (IV) corresponding to rewiring performed in both the 
layers as compared to that of the case (III), which corresponds to rewiring  performed in only
one layer. However, further increments 
in $D_{x}$ as well as in $E_{y}$ (as $E_y = D_x$ for $D_x > 1$) values lead to a 
domination of the contribution of stronger couplings in 
$L^{AV}$ and as a result, the efficiency for case (IV) converges towards that of the case (III).

Figure.~\ref{Case} (b) depicts that efficiency of the optimization is same  
for the cases (V) and (VI), although there are huge differences in the computational
cost for the optimization process. Case (V) considers rewiring performed in
both layers and case (VI) has only one layer being rewired.  
Equation~(\ref{EQ2}) explains this behaviour 
since for both cases the $R$ values depend on $\lambda_{max}$ which is only determined 
by the layer having 
the stronger intra-layer coupling strength going through rewiring for
both the cases. Finally, we find that the results 
about
the efficiency for cases (III) and (IV) are valid for the denser networks as well ( Fig.\ref{Case}~(c)). 
The one difference as compared  to the sparser networks is that the efficiency is equal 
for both cases having larger values of $D_{x}$. Again, this behaviour
arises due to the nature of Eq.~(\ref{EQ3}), an equation that becomes
more accurate for larger values of  $D_{x}$ i.e., for the multiplex networks 
having stronger inter-layer couplings.

\begin{figure}[t]
\centerline{\includegraphics[width=0.9\columnwidth,height=4cm]{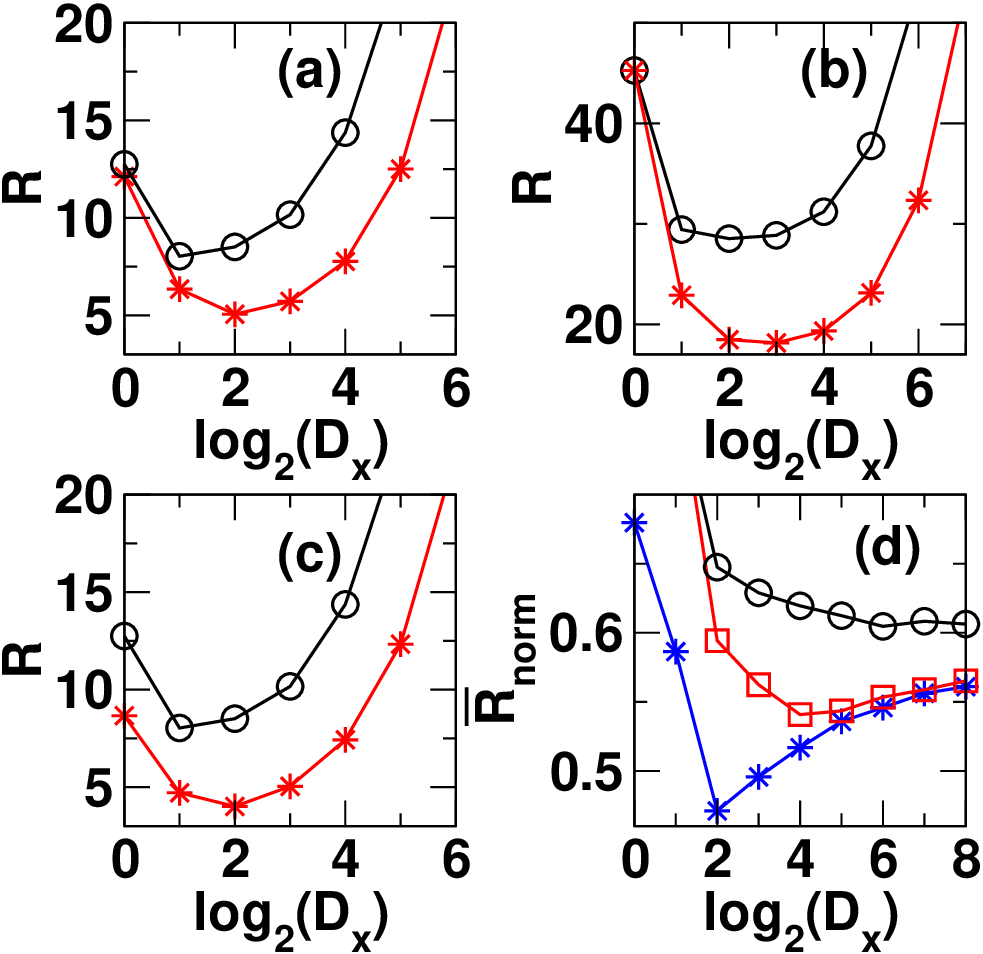}}
\caption{The initial (circles) and optimized values (star) of
the eigenvalue ratio ($R$)
with an increase in the inter-layer coupling strength $D_{x}$ for (a) one layer having
a fixed ER configuration and the other layer is rewired, (b) 
the fixed layer represented by a SF network and the other layer is 
rewired, (c) both layers are initially represented
by ER networks and both layers are rewired during the optimization.
(d) depicts the efficiency of optimization ($R_{\mathrm{norm}}$) when the fixed layer 
is represented by an ER network and the other layer is rewired (square). The case
when the fixed layer is represented by the SF configuration and 
the other layer is rewired is depicted by circles. The case when both layers 
are represented by ER networks before optimization and both the 
layers are rewired during the optimization is depicted by star. For all the cases, network size in each layer is 
500 with average degree 10.}
\label{FixStru}
\end{figure}
To study the dependence of the optimization process on the topology of 
one fixed layer, we consider the initial fixed layer constructed
by the Watts-Strogatz model with various rewiring probabilities
$p_r$.  The small-world transition (Fig.~\ref{SWtr}(a)) 
for the Watts-Strogatz model is characterised by a 
clustering coefficient as high as that of the regular network
and the characteristics path length being as small as that of 
the random networks. 
For an ER network representing the layer going through the rewiring
during the optimization process, and for small values of $p_r$ 
typically smaller than
the SW transition, the initial and
the optimized multiplex networks have both the same
synchronizability (Fig. 2(b)). For $p_r$ larger than the value for the SW transition,
synchronizability of both the initial and the optimized multiplex networks
start increasing and attains its maximum value
(the lowest $R$ value) at a rewiring probability which is much higher 
than the critical parameter for the SW transition $p_r$, but much smaller than $p_r=1$. Such a dependence of synchronizability on $p_r$ is the result of an
interplay between the degree homogeneity of the fixed layer and the layer going through the optimization.  
Initially for a $p_r$ being smaller than the value for the SW 
transition, the diameter of the fixed layer is large 
resulting in a poor synchronizability of the entire multiplex network.
For $p_r$ being greater than the value for the SW transition, as long 
as the fixed layer has still small degree heterogeneity, the optimized multiplex networks 
possess the following topological characteristics 
contributing to better synchronizability; (1)
degree homogeneity for both the fixed layer and the layer experiencing the
rewiring (i.e., the distribution of degrees is not broad), (2) small values of both 
the average path length and the diameter of the entire multiplex 
networks. 
For the fixed layer generated with $p_r=1$ or close to 1, 
though the diameter and the average path length of the entire network
are still small, the degree heterogeneity of the fixed layer
is high enough which does not get balanced by the 
rewiring of another layer during the optimization process, resulting in a smaller 
synchronizability of
the optimized network. The value of $p_r$, corresponding to the maximally synchronizable
network achieved through the evolution process, decreases as the average degree of the 
initial networks increases. This shift in $p_r$ towards the lower 
values arises due to the fact that for denser networks, 
even very small rewiring probability values are sufficient to destroy the degree homogeneity of the initial 
fixed layer, having a similar impact on the synchronizability of the
final evolved network.

Moreover, optimization of denser networks
leads to a less synchronizable evolved networks than those achieved by optimizing
sparser networks, since denser networks possess a larger amount of 
mismatch in the inter and the intra-layer
connections \cite{Liang}.  
For the sparser networks, the efficiency of synchronizability 
is high for a very large range of $p_r$. However, denser networks reflect 
comparatively a lesser efficiency of the optimization, i.e., smaller values of
$R_{norm}$ (Fig~\ref{SWtr}(c)), as the fixed layer 
restricts the value of $R$ to decrease
beyond a limit even though the second layer is rewired to enhance the
synchronizability of the entire multiplex network.

Further, to study the impact of change in the structural properties of the fixed layer
on the efficiency of optimization, we consider the fixed layer being represented by
ER random and scale-free networks. Fig.~\ref{FixStru}(a) depicts that there is a decrease
in $R$ with an initial increase in $D_{x}$. With a further increase in
$D_x$, $R$ starts increasing for the case of ER representing the fixed layer.  
For the fixed layer being represented by a scale-free network, $R$ first decreases with an 
initial increase in the value of
$D_{x}$, and after attaining a minimum value it remains almost constant for a further increase
in $D_{x}$ or for larger $D_x$ values.  As $D_x$ increases further,
$R$ finally starts increasing.  Again, similar to the previous case of fixed layer represented by
ER network, the networks with lower $D_x$ values are not optimizable (Fig.~\ref{FixStru}(b)).  This result is in contrast to
the behaviour exhibited for the un-restricted rewiring case. When both the layers are 
rewired, the networks are optimizable for all the $D_{x}$ values (Fig \ref{FixStru}(c)). 
Fig.~\ref{FixStru}(d) reflects that for the unrestricted rewiring, i.e. for rewiring taking place
in both the layers, the efficiency of optimization is maximum for a certain value of
$D_{x}$ after which it again decreases.
Interestingly, $D_{x}$ for which efficiency is maximum
is shifted towards a larger value for the case of fixed layer being
represented by ER random networks which also corresponds to the  
maximum efficiency. There is more shift towards a larger value for the case of fixed layer represented
by the SF networks. The reason behind this shift is that the local minima of $R$ gets shifted towards a
higher value of $D_x$ for the layer having the scale-free architecture [12].

{\it Conclusion:}
Our results show that there are several pathways to improve synchronizability of multiplex networks, either by altering parameters such as those that promote integration of the layers (increasing the inter-layer coupling strength), or by evolving the network topology by rewiring edges within layers,  under an optimization process. The surprising result is however that optimization of a single layer can achieve networks that are roughly as capable to synchronize as networks where all the layers are evolved under similar optimization criteria. This result is particularly relevant to works intended to improve synchronization of systems where only one layer is accessible or when one wants to optimize a system in a very cost effective fashion. Having in mind that 
real-world systems are very large, complex, and composed by many layers, our work points that optimization in such systems can indeed be carried out.           

We have also studied the effectiveness of the optimization process, measured 
by the network synchronizability achieved through the evolution process, 
when the initial pre-evolved networks have different initial topologies. 
We found that the optimization leads to the maximum synchronizable multiplex 
networks when the fixed non-evolved layer has a topology lying in between a 
network with incipient small-world and fully random topologies.  

Networks theory has proven its aptness in providing insights into controllability 
at a fundamental level. The controllability is desirable for dynamical behavior associated 
with the functionality of real-world systems. In traditional approaches, external inputs are 
imposed to affect the dynamics of few nodes which further causes a control of 
the entire system \cite{control}. Our work might refine the concept of 
controllability  
by addition of a new system (one layer) that changes the dynamical evolution of 
the entire system (multiplex) to a 
desired behavior.
Further, our work might complements works on controllability by creating more
synchronous evolved networks  that could be more controllable.

{\it Acknowledgements:} SJ acknowledges DST grants EMR/2014/000368/PHY
and EMR/2016/001921/PHY for financial support.

\end{document}